# HI-HCQC: A Tightly-Coupled Hardware Interface with High-Efficiency Communication for Hybrid Classical-Quantum Computing

Shibo Liang[1,†], Junchao Wang[1, †,*], Zeyuan Wang[1], Feng Wang[1], Xiaoyu Li[2], Lei Li[1], FuDong Liu[1], Zheng Shan[1,#]

[1]State Key Laboratory of Mathematical Engineering and Advanced Computing, Zhengzhou 450002, China

[2]University Of Electronic Science And Technology Of China

[†]These authors contributed equally to this work

[*]Corresponding author.Email: wangjunchao11@126.com

[#]Corresponding author.Email: shanzhengzz@163.com

## Abstract

Hybrid classical-quantum computing requires frequent data exchange between classical processors and quantum control hardware. However, existing superconducting quantum control systems are commonly connected through loosely coupled interfaces such as Ethernet, resulting in high communication latency and limited task throughput. To address this issue, we present HI-HCQC, an RFSoC-based hardware interface for tightly coupled hybrid classical-quantum computing. HI-HCQC integrates high-speed RF-DACs, RF-ADCs, programmable logic, embedded processors, clock synchronization circuits, and a PCIe Gen3 x8 interface, enabling direct microwave pulse synthesis, qubit readout, and high-throughput data transfer between host servers and quantum measurement-control units. Experimental results show that HI-HCQC supports six control channels and one multiplexed readout channel, achieves stable microwave generation and acquisition, and successfully performs qubit spectroscopy, Rabi oscillation, T1 measurement, single-shot readout, randomized benchmarking, and CZ-gate characterization. Compared with a conventional control system, HI-HCQC reduces end-to-end execution latency for representative quantum gate and circuit tasks and significantly improves task throughput. These results demonstrate that PCIe-coupled RFSoC control hardware provides a practical foundation for scalable and efficient hybrid classical-quantum computing systems.

## Introduction

In recent years, quantum computing has garnered significant attention in the forefront of scientific research. Compared to traditional classical computers, quantum computing theoretically possesses the potential for exponential acceleration in certain computational tasks, particularly in fields such as integer factorization[1], quantum chemistry[2], and machine learning[3]. Some research institutions are actively engaged in the development of quantum computing technologies, with the primary physical implementations including superconducting qubits[4], trapped ions[5], and photonic quantum systems[6]. Among these, superconducting quantum chips, due to their compatibility and similarity with current integrated circuit technologies, have rapidly evolved as one of the most promising candidates for universal quantum computing. Meanwhile the growth of artificial intelligence demands increased computational power, highlighting the limitations of single-architecture processors in handling complex data processing tasks. Quantum computing, despite its prowess in specific areas, faces challenges in problem diversity and error rates, preventing it from fully replacing conventional high-performance computing (HPC). Consequently, the synergy between HPC and quantum computing within hybrid architectures is crucial for advancing computational capabilities.

Superconducting quantum chips rely on circuit quantum electrodynamics (cQED)[7], which integrates microwave resonators and superconducting qubits through microwave control techniques. As we are staying in the Noisy Intermediate-Scale Quantum (NISQ) era[8], there is an urgent need for improved qubit control and measurement hardware to manage the increasing complexity of quantum processors. Moreover, quantum computing devices and classical computing devices should be closely coupled. For example, the quantum errors can be mitigated assisted with the classical approach such as error mitigation[44]. A complete application should also integrate the quantum computing and classical computing. For example, the variational quantum algorithm like QAOA and VQE requires the parameters updated with the classical computing. This needs to address developing high performance, scalable and compact control systems without compromising on precision and functionality.

In addressing these challenges, we design HI-HCQC, a control system for superconducting quantum computers implemented on an FPGA board. Then we propose a hybrid computing system integrating key components such as classical servers, HI-HCQC, and quantum computers. Utilizing the Xilinx Zynq Ultrascale+ RFSoC[9], the HI-HCQC incorporates high-speed Radio Frequency-Digital to Analog Converter (RF-DACs), high-speed Radio Frequency-Digital to Analog Converter (RF-ADCs), programmable logic (PL), processing system (PS), and high-speed data transmission interfaces, enabling precise microwave signal and pulse synthesis and measurement. This architecture enhances the interoperability between classical and quantum computing, offering a streamlined solution for qubit control and improving the collaborative efficiency of hybrid computing systems. The system's effectiveness is demonstrated through extensive qubit calibration experiments.

The main contributions of this work are as follows:

- System Architecture: We design an RFSoC-based superconducting quantum control and readout board with six RF control channels, one multiplexed readout channel, and integrated clock synchronization.

- High-Speed Interconnect: We implement a PCIe Gen3 x8-based data path with DMA support, enabling tighter coupling between classical servers and quantum measurement-control hardware, which reduces the data transmission latency by a factor of 169 compared to traditional Ethernet-based systems.

- Comprehensive Evaluation: We build a hybrid classical-quantum system architecture and experimentally validate it through single-qubit calibration, single-shot readout, randomized benchmarking, and CZ-gate characterization. System-level evaluations demonstrate an improvement of overall quantum gate throughput by over 320 times.

## Control System of Superconducting Quantum Computers

The operational foundation of superconducting qubits derives from the quantum mechanical characteristics inherent in superconducting circuits[10]. These qubits are delineated by a Hamiltonian, a mathematical representation of the total energy of the system. In quantum mechanics, the Hamiltonian functions as the operator for total energy, encapsulating the dynamic behavior of a given system. In the specific realm of superconducting qubits, particularly those utilizing Josephson Junctions, the Hamiltonian assumes an expression as follows:

$$H = \frac{Q^2}{2C} - E_J \cos(\varphi) \quad (1)$$

where Q represents the charge, C the capacitance, $E_J$ the Josephson junction energy, and φ the superconducting phase difference.

Within superconducting quantum computers, the control of qubits is directly manipulated by radio-frequency pulses generated through microwaves, as illustrated in Fig. 1. The execution of quantum gates, fundamental operations in quantum computing, involves modifying parameters within the Hamiltonian—such as frequency, amplitude, or phase of the microwave pulses. These alterations systematically transform the quantum states of qubits.

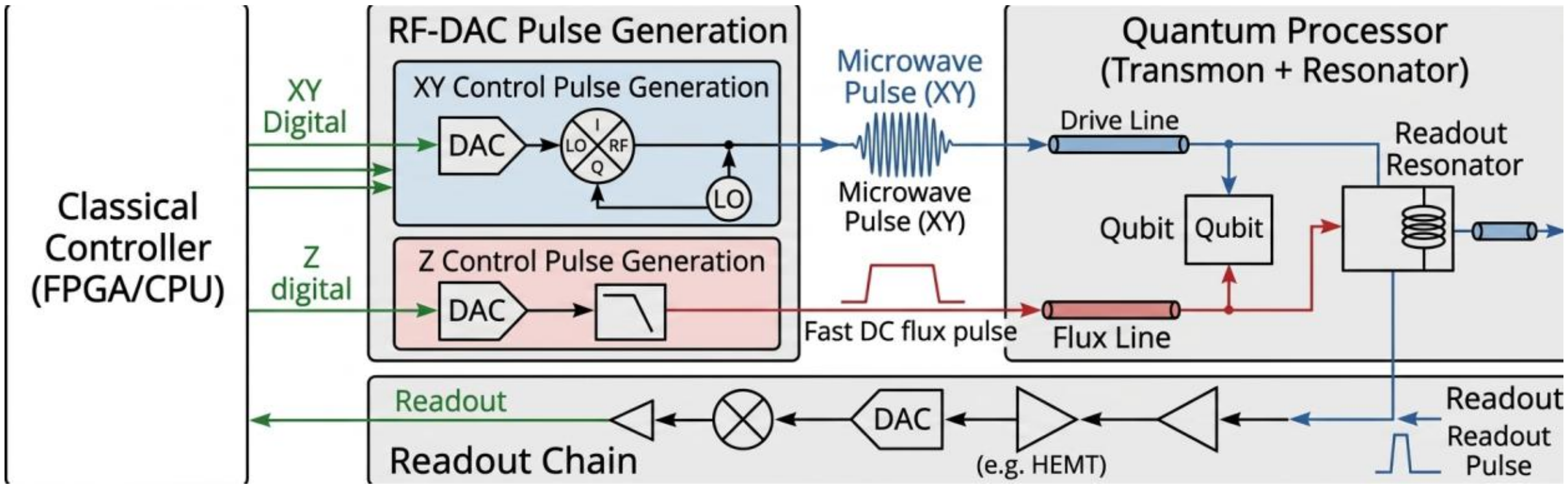

**Figure 1.** Control principles of superconducting qubits.

One prevalent type of superconducting qubit is the Transmon qubit[45], wherein transitions between energy levels are induced through precisely orchestrated microwave pulses. By skillfully applying these pulses, fundamental quantum gates, including Pauli-X, Pauli-Y, Pauli-Z, Hadamard, and controlled-NOT (CNOT) gates, can be instantiated[11]-[13]. These gates serve as foundational elements for quantum algorithms, enabling qubits to engage in intricate computational processes. To illustrate, the implementation of a X gate (bit-flip gate) can be exemplified through the modeling of the pulse's Hamiltonian as follows:

$$\mathrm{H}_{\mathrm{pulse}} = \Omega\cos(\omega t + \phi)\sigma_x \quad (2)$$

where Ω is the amplitude, ω the frequency, ϕ the phase, and $\sigma_x$ a Pauli matrix.

The resonance between the pulse and the qubit's energy level difference facilitates the quantum state transition. The parameters Ω and the pulse duration t are tuned to achieve a transition from the ground state $|0\rangle$ to the excited state $|1\rangle$, representing the X gate operation. The condition for a complete state flip is met when $\Omega t = \pi$.

The coherence and fidelity of these operations are key challenges in superconducting quantum computing, as qubits can lose their quantum state due to interaction with the environment (decoherence). Therefore, most of the research in this field focus on improving qubit design and control techniques to maintain coherence for longer periods and execute gates with high precision[14],[15].

In the initial phases of experimentation with superconducting qubits, conventional practices involved the utilization of costly general-purpose test equipment or proprietary embedded systems[16]-[18]. Presently, the prevailing trend in control and readout systems for superconducting quantum computers leans towards reliance on commercial devices, rather than bespoke designs tailored for the specific task at hand. This has resulted in the integration of multiple components, contributing to escalated costs and increased system complexity. Furthermore, the intrinsic limitations in qubit lifespan, coupled with the response time of classical electronics, impose constraints on quantum computer performance. The efficacy of control systems is intricately linked to their turnaround time, and the modular configuration of extensive systems introduces extended turnaround times.

The current superconducting quantum measurement and control system mainly includes three key components, as shown in Fig. 2. The host computer, as a pivotal component, assumes the responsibility of receiving user commands and formulating measurement and control tasks. And the microwave control system comprises room-temperature measurement and control equipment integrated with a low-temperature line system[19],[20].Operating within a low-temperature environment approaching absolute zero, superconducting quantum chips rely on a dilution refrigerator to provide requisite conditions, including low temperature, vacuum, and shielded magnetic fields.

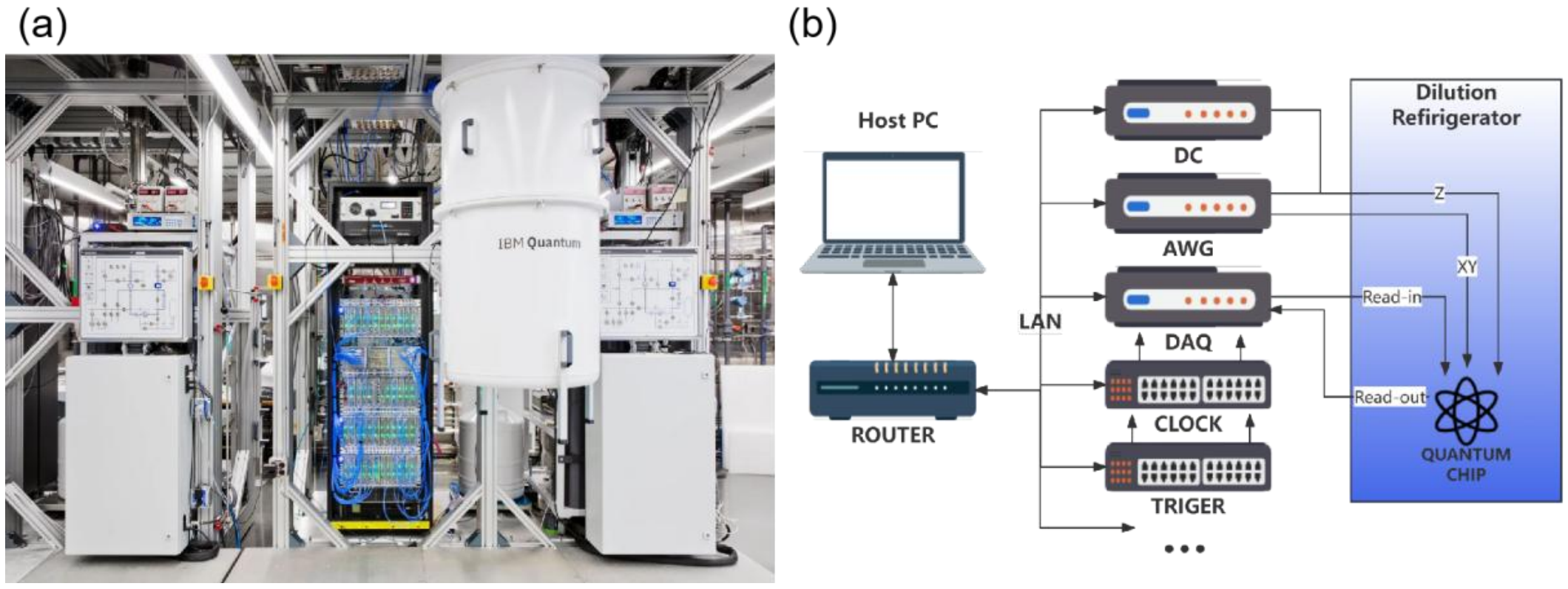

**Figure 2.** Subfigure (a) is a superconducting quantum computer system of IBM. Subfigure (b) is the architecture diagram of quantum computing system.

As we are staying in the Noisy Intermediate-Scale Quantum (NISQ) era[8], there is an urgent need for improved qubit control and measurement hardware to manage the increasing complexity of quantum processors. Moreover, quantum computing devices and classical computing devices should be closely coupled. For example, the quantum errors can be mitigated assisted with the classical approach such as error mitigation[44]. A complete application should also integrate the quantum computing and classical computing. For example, the variational quantum algorithm like QAOA and VQE requires the parameters updated with the classical computing. This needs to address developing high performance, scalable and compact control systems without compromising on precision and functionality.

In addressing these challenges, we design HI-HCQC, a control system for superconducting quantum computers implemented on an FPGA board. Then we propose a hybrid computing system integrating key components such as classical servers, HI-HCQC, and quantum computers. Utilizing the Xilinx Zynq Ultrascale+ RFSoC[9], the HI-HCQC incorporates high-speed Radio Frequency-Digital to Analog Converter (RF-DACs), high-speed Radio Frequency-Digital to Analog Converter (RF-ADCs), programmable logic (PL), processing system (PS), and high-speed data transmission interfaces, enabling precise microwave signal and pulse synthesis and measurement. This architecture enhances the interoperability between classical and quantum computing, offering a streamlined solution for qubit control and improving the collaborative efficiency of hybrid computing systems. The system's effectiveness is demonstrated through extensive qubit calibration experiments.

As the landscape of superconducting quantum computers evolves, there is a burgeoning demand for more sophisticated and specialized devices. In response, certain platforms are adopting integrated approaches that amalgamate DACs, ADCs, and FPGAs, facilitating real-time control and measurement of quantum computers[21]-[28]. These controllers primarily employ the traditional In-phase and Quadrature (IQ) mixing method, synthesizing intermediate-frequency signals and executing up-conversion. As shown in Fig. 3, renowned vendors such as BBN[29], Keysight[30], Zurich Instruments[31], and Quantum Machines[32] have introduced commercial products endowed with FPGA-enabled real-time pulse synthesis and readout capabilities.

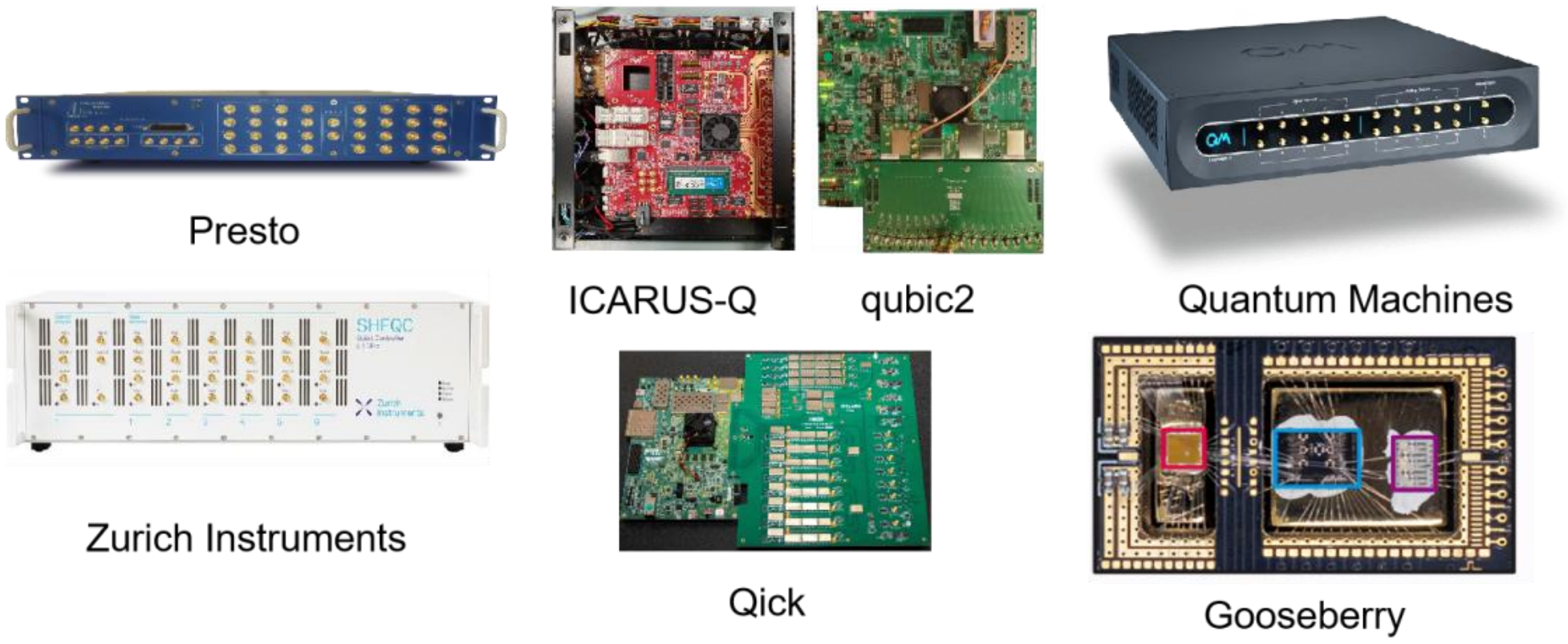

**Figure 3.** Some Control Systems of Superconducting Quantum Computers.

Recent advancements in RF-DAC technology, coupled with a substantial increase in sampling rates, have ushered in the capability to directly synthesize microwave pulses without the necessity for an up-conversion step. This breakthrough streamlines the system calibration process, enabling direct digital synthesis (DDS) of RF waveforms. This approach significantly diminishes the hardware footprint and energy consumption of control systems, with validation and application in various academic laboratories such as qubic2.0[33],Presto[34],Qick[35],SQ-CARS[36] and ICARUS-Q[37].As shown in Tab. 1, we compared the HI-HCQC with some Quantum Control Systems. These systems use different platforms and chips. And all of them use Ethernet interfaces except our work.

| | QICK | ICARUS-Q | Presto | SQ-CARS | Qubic2.0 | HI-HCQC |
|---|---|---|---|---|---|---|
| Platform | ZCU111 | HTG-ZRF16 | ZCU208/ ZCU216 | ZCU111 | ZCU216 | CC305 |
| RFSoC | XCZU28DR | XCZU29DR | XCZU49DR | XCZU28DR | XCZU49DR | XCZU47DR |
| DAC(GSPS) | 6.144 | 6.144 | 10 | 6.144 | 9.85 | 8 |
| ADC(GSPS) | 4.096 | 1.96608 | 5 | 4.096 | 2.5 | 4 |
| Time Synchronization | tProcessor | MTS | MTS | MTS | MTS | MTS |
| Hardware Interface | Ethernet | Ethernet | Ethernet | Ethernet | Ethernet | PCIe |

**Table 1.** Comparison of HI-HCQC with various Quantum Control Systems.

Furthermore, significant advancements have been achieved in the field of low-temperature measurement and control systems, exemplified by the development of technologies like Gooseberry[38]. The emergence and adoption of open-source quantum measurement and control systems represent a crucial trend. As technological progress persists, these systems are poised to become more efficient and intelligent, adapting to the evolving computational needs of the future and instigating revolutionary changes in the field of quantum computing.

## Architecture of HI-HCQC

HI-HCQC embraces a modular design characterized by hardware and firmware scalability, effectively accommodating the increasing number of qubits. Due to limitations in board size and technological constraints, the present configuration extends support to 6-way XY control and 1-way readout, facilitating comprehensive governance over 6 qubits. The prevailing architecture of superconducting quantum chips gravitates towards frequency multiplexing for quantum bit readout, albeit necessitating a solitary microwave line for quantum bit control[39]. In the RFSoC setup, a surplus of DAC channels compared to ADC channels typically proves imperative, emblematic of the intricate nature and requisites inherent to quantum computing tasks. HI-HCQC can be seamlessly expanded to cater to the control and readout of an augmented number of qubits, either by increasing the number of DACs and ADCs or by harnessing frequency multiplexing capabilities within the available channels.

The system architecture consists of two integral components: the PS and the PL. The PS, driven by a multi-core ARM processor running the Linux operating system, assumes the responsibility for high-level control and data management functions. Utilizing the PYNQ[40] library and drivers, it facilitates Direct Memory Access (DMA) to optimize data transmission. End users are granted access through Jupyter notebooks, streamlining the definition and management of quantum measurement and control experiments. Empowered by the advanced computational capabilities of the PS, the system is well-equipped to execute intricate data analysis and learning algorithms, thereby supporting activities such as quantum state reconstruction, noise analysis, and system calibration.

In typical quantum experiments, as depicted in Fig. 4, the PS dispatches instructions to the PL, where the signal generator module generates control pulses that influence the quantum bits. The state of the quantum bits post-interaction is captured by the high-speed ADC within the PL. Subsequently, the data undergoes digital downconversion and filtering, with the processed readout data transmitted back to the PS for analysis. The FPGA assumes a pivotal role in ensuring precise timing for quantum operations, managing the data flow within the PL, and orchestrating the synchronization of signal generation, readout, and data retrieval.

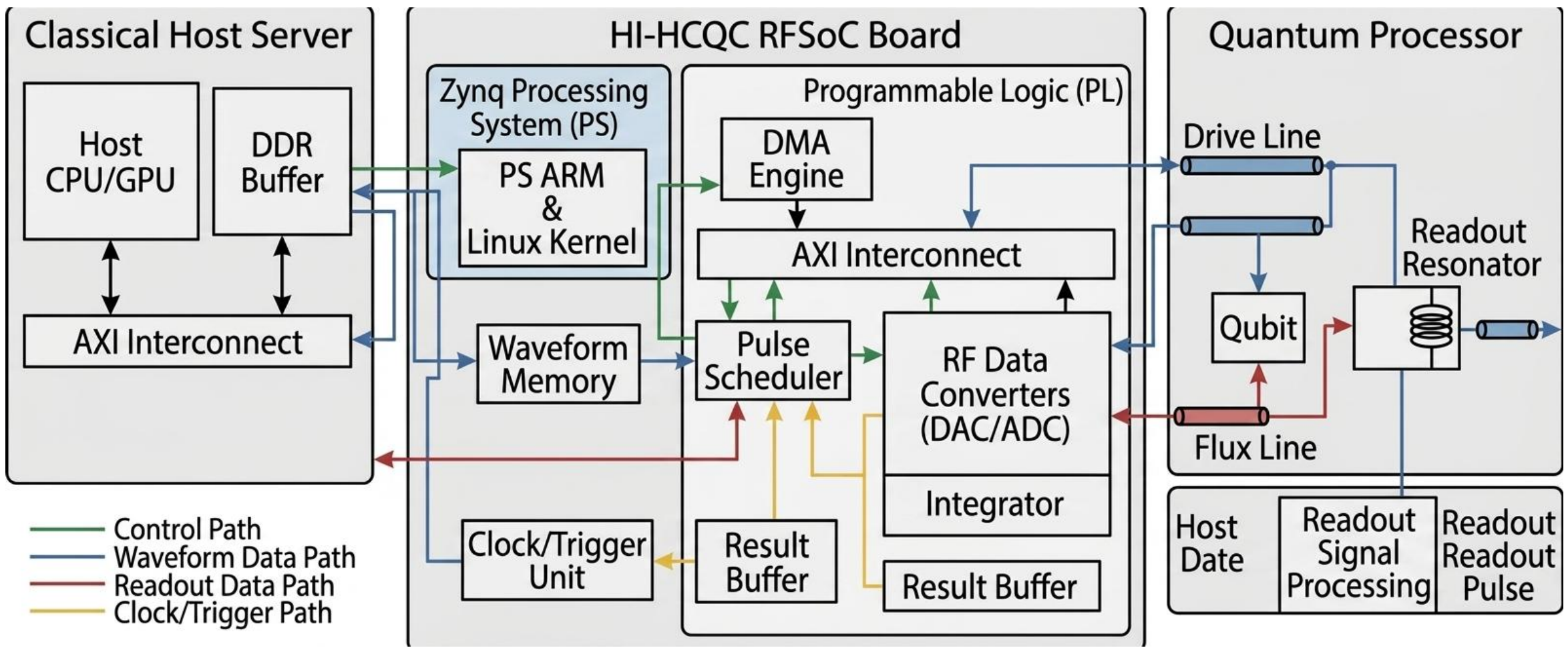

**Figure 4.** Architecture of HI-HCQC.

HI-HCQC incorporates an internal clock management circuit encompassing clock synthesis, damping, and routing components, thereby affording flexible clock distribution. Furthermore, it extends support for external reference clock and trigger signal input, ensuring synchronized clock operation across multiple boards. In terms of data transmission, the board leverages a PCIE x8 GEN3 interface boasting a maximum rate of 8GB/s, thus establishing a dependable and high-performance data transmission conduit. This can be synergistically employed with other devices to achieve a more condensed experimental layout.

Additionally, HI-HCQC is equipped with auxiliary components tailored for signal conditioning and system expansion. The clock unit dispenses a high-precision and stable clock reference, facilitating the direct generation of a low-phase-noise, stable 100MHz reference clock signal, thus accommodating the system-level expansion of multiple measurements and control units. The trigger unit yields highly synchronized, adjustable delay trigger output signals, effecting synchronous operations across multiple measurement and control units. These auxiliary devices are instrumental for cascading the expansion of multiple discrete units.

**Hardware Interface**

We adopt a PCIe half-length hardware standard that is compatible with classical servers. This compatibility allows for direct physical integration into classical servers, enabling the seamless integration of quantum measurement and control units with classical computing units. The utilization of a high-speed PCIe interface for interconnection with classical computing units leads to a substantial reduction in interconnection latency.

As shown in Fig. 5, the PCIe Interface for RFSoC Control and Management involves the use of a DMA AXI-Lite interface for RFDC and other IP control and status. The HI-HCQC incorporates a general-purpose register file for various control, status, and statistics counters. Access to these registers is achieved through a 32-bit wide interface, facilitated by the DMA driver via the PCIe interface. The Data Management Architecture comprises a Xilinx DMA subsystem responsible for DMA transfers between the Host PC and the board. These transfers, along with the host interface, are facilitated through the PCIe interface on the Host PC.

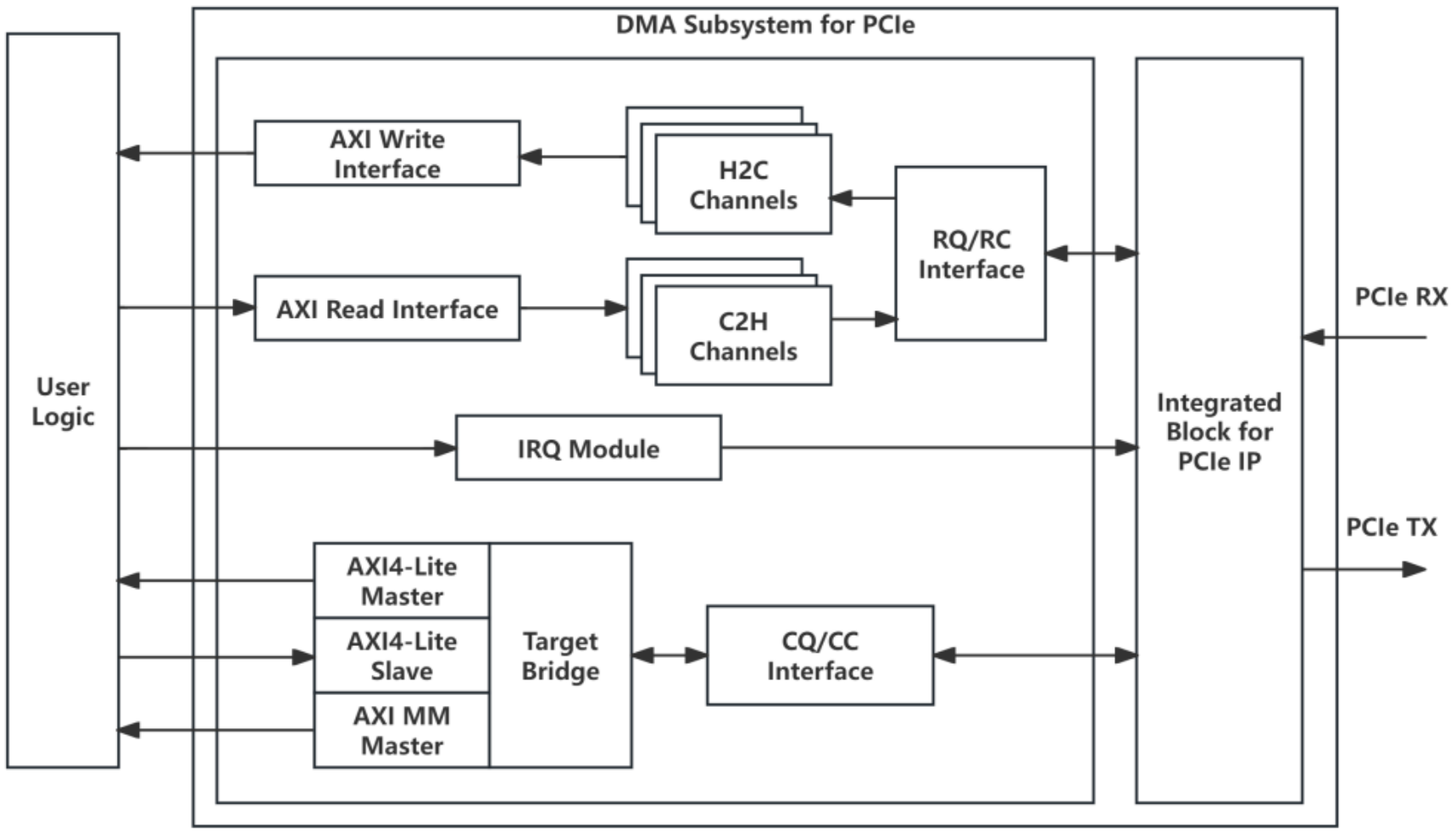

**Figure 5.** DMA Subsystem for PCIe.

### Microwave Generation

In consideration of critical factors such as signal quality, channel density, and single-channel cost, we have opted for the direct output method of RF-DAC in the Nyquist zone (NZ)[26]. This approach allows for the generation of multi-channel high-fidelity microwaves by integrating RF data converters and DDS functions. The measurement and control board can directly generate and modulate RF signals, thereby achieving direct digital synthesis. This eliminates undesirable transitions driven by mixer spurs and obviates the need for meticulous calibration of IQ mixer offsets and gains. As a result, the system structure is simplified, and overall performance is enhanced.

As depicted in Fig. 6, the direct output solution of RF DAC in the NZ delivers a stable output signal, reducing the necessity for frequent calibration. However, it is essential to acknowledge that this approach may lead to phenomena across the NZ, necessitating enhancements in adaptability to diverse system requirements.

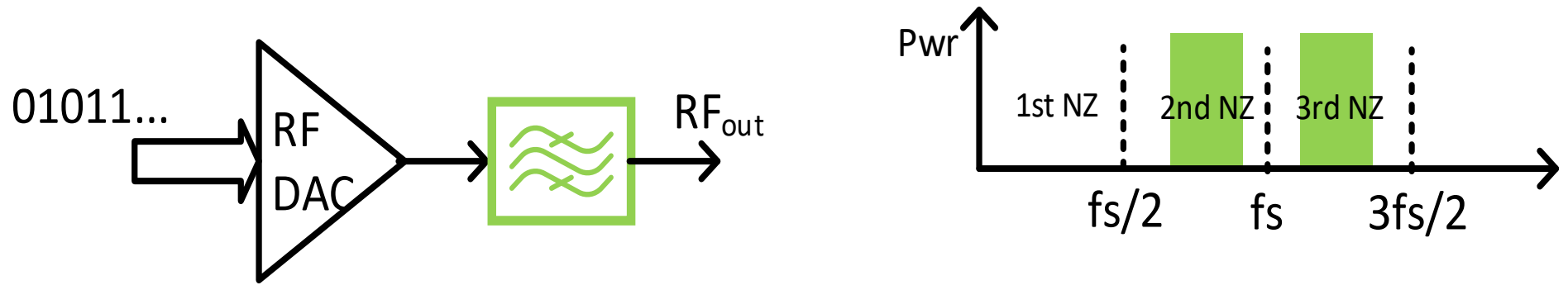

**Figure 6.** The direct output method of RF-DAC in the NZ.

For pulse generation, digital IQ mixers and Numerically Controlled Oscillators (NCOs) are employed for signal interpolation and up-conversion. The system supports parameter updates between pulses, encompassing amplitude, phase, and frequency adjustments. Each quantum bit is allocated a dedicated DAC channel. In terms of signal measurement, digital IQ mixers and NCOs down-convert the signal to baseband, after which the FPGA's parallel processing capabilities are harnessed for analysis. This facilitates the rapid execution of intricate data processing tasks, including real-time data compression, signal filtering, and error correction. The digitization of qubit readout signals is executed by the ADC, followed by mixing with a digital local oscillator, integration, and subsequent storage for quantum bit state discrimination. The system also supports parameterization and triggering of RF control/measurement pulses.

## Hybrid Classical-Quantum Computing System

To delve into the intricacies of algorithms and foster the development of applications that harness quantum advantages, it is imperative to employ technical methodologies that facilitate dynamic workflows across disparate system architectures. The paper delineates an integrated approach, as depicted in Fig. 7, that amalgamates Quantum Processing Units (QPUs), GPUs, and CPUs within a hybrid computing framework.

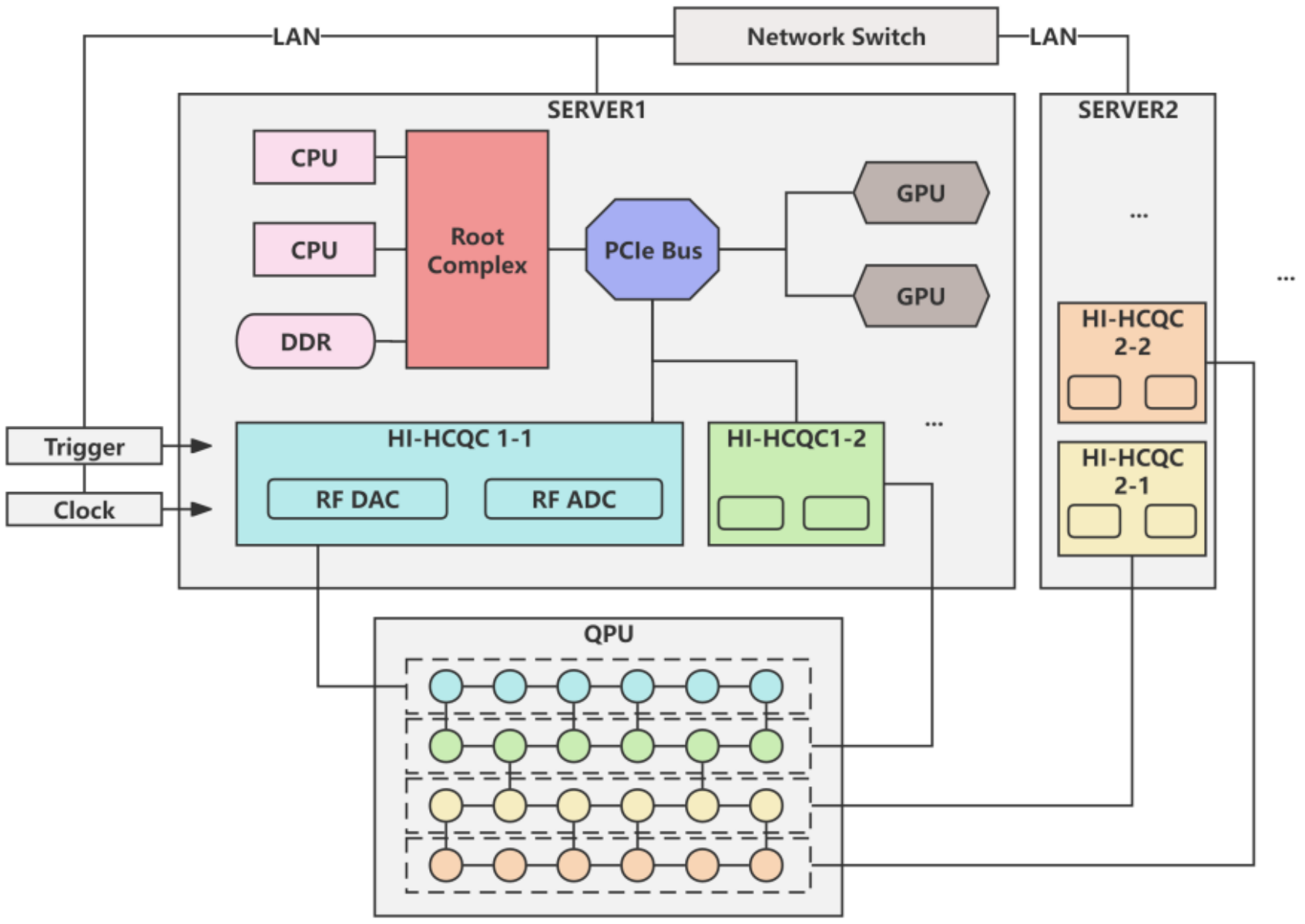

**Figure 7.** Architecture of Hybrid Classical-Quantum Computing System.

This hybrid computing system represents a fusion of classical and quantum computing technologies, engineered to tackle intricate computational challenges by capitalizing on the unique strengths inherent to each computing paradigm. At the foundation of this system lies a network of multiple servers, linked via a Local Area Network (LAN), which forms a robust classical computing infrastructure. Each server within this network is equipped with GPUs and dedicated quantum measurement and control apparatus, all of which are interconnected through the high-speed PCIe bus, boasting a data transmission bandwidth of up to 8 GB/s. This high bandwidth is instrumental in enhancing the efficiency of waveform loading and the transmission of readout data. The system's architecture allows for scalable and adaptable hardware resource allocation, contingent upon the computational demands and the availability of PCIe slots.

For endeavors in large-scale quantum computing, particularly those involving the manipulation of hundreds of qubits, the system adopts a distributed architecture. This configuration encompasses multiple servers, each outfitted with an array of HI-HCQC operating in concert. Such a distributed and collaborative framework is essential for the parallel and synchronized control of a vast number of qubits, a prerequisite for executing complex quantum computations effectively.

The scalability of the system is often constrained by the channel capacity of individual FPGAs or RFSoCs, necessitating the synchronization of multiple devices to enhance system efficacy. A critical aspect of this synchronization is the precise phase alignment of clock signals across all modules, which is paramount for ensuring the coherent synchronization of DAC and ADC signals, along with the precise alignment of clock counters. To achieve this, high-precision clocks are implemented across the HI-HCQC, thereby establishing a uniform time reference throughout the system. At the heart of our synchronization strategy is the deployment of a high-stability master clock. A specialized clock distribution amplifier, which utilizes differential signaling, is employed to distribute the clock signal to each RFSoC board while preserving signal integrity and minimizing noise interference. Each RFSoC board is outfitted with a Phase-Locked Loop (PLL) circuit, specifically chosen for its low jitter properties, to synchronize the board's local oscillator with the master clock. Considerable attention has been devoted to the selection of cabling and the design of the layout, with a focus on utilizing high-quality, impedance-matched cables and maintaining uniform cable lengths to preclude timing discrepancies.

The design philosophy of the HI-HCQC backend, grounded in the Qiskit[41] architecture, is engineered to facilitate an efficient interface between quantum computing hardware and software. We adopt a highly abstracted framework, enabling the execution of identical quantum algorithms across diverse hardware platforms. At the foundational level, the HI-HCQC interface layer is tasked with direct hardware interaction, ensuring accurate waveform data playback and result collection. Data processing is bifurcated into initial transformation to conform to the Qiskit format, followed by detailed analysis via the Qiskit data processor.

This paper has realized an internal hardware layout scheme for server complete machines oriented to classical-quantum heterogeneous collaborative computing scenarios, as illustrated in Appendix A-1.

# Performance Characterisation

### A.Bench Test

The HI-HCQC features six control channels and one readout channel, employing a direct RF output design. This DAC features a high sampling rate of 8Gsps and a sampling precision of 14 bits, enabling it to provide output signals within the frequency range of 4.2~5.5 GHz. It offers a wide instantaneous bandwidth exceeding 1GHz and a minimum pulse width of 30 microseconds. The output power of the DAC is adjustable with a maximum attenuation of 30dB and a step size of 0.25 dB. Additionally, it boasts an SFDR performance superior to -45dBc, indicating minimal harmonic distortions, and phase noise better than -80dBc/Hz @ 1kHz, 4.8 GHz, demonstrating excellent phase stability at specific frequencies. Lastly, the DAC possesses an RF channel impedance of 50 Ω, making it suitable for connecting with other RF devices or antennas. The ADC is featuring a maximum sampling rate of 4Gsps and a sampling precision of 14 bits. It has an input frequency range of 6.0~7.5GHz and can handle full-scale input power levels up to ⩽10dBm (minimum power -50dBm) with a precision of 1dB. The demodulation analysis pulse width is ⩾10us, and it can analyze at least 4 Qubits per channel simultaneously with a minimum bandwidth of 1GHz.

For the performance of the board, we tested its pulse width, power, among other metrics. The test results, as shown in Tab. 2, met the standard levels, indicating that the board can effectively support the transmission and reception of microwave data and accurately generate the required RF pulses for controlling and measuring quantum bits.

| | Control channel（DACx6） | | | | |
|---|---|---|---|---|---|
| | Pulse width | Power | Stray | Phase noise | Gausssian pulse |
| result | 30us | -8.72dBm | -65dBc | -97dBc/Hz@1kHz,4.8 GHz | 12us |
| | Read channel（ADC） | | | | |
| | Pulse width | Power | Stray | Phase noise | Channel isolation |
| result | 10us | -11.7dBm | -56dBc | -93dBc/Hz@1kHz,4.8 GHz | 80.2dBc |

**Table 2.** Channel parameters of the HI-HCQC.

### B. Characterization of a Transmon qubit

In quantum computing, the calibration of superconducting qubits is essential to ensure their accurate and reliable operation. This paper conducts a series of qubit calibration experiments on a sample of a fixed-frequency superconducting qubit to validate the measurement and control functionality of the hybrid computing system. The qubit was mounted to a dilution refrigerator with various attenuators and filters on the input microwave lines.

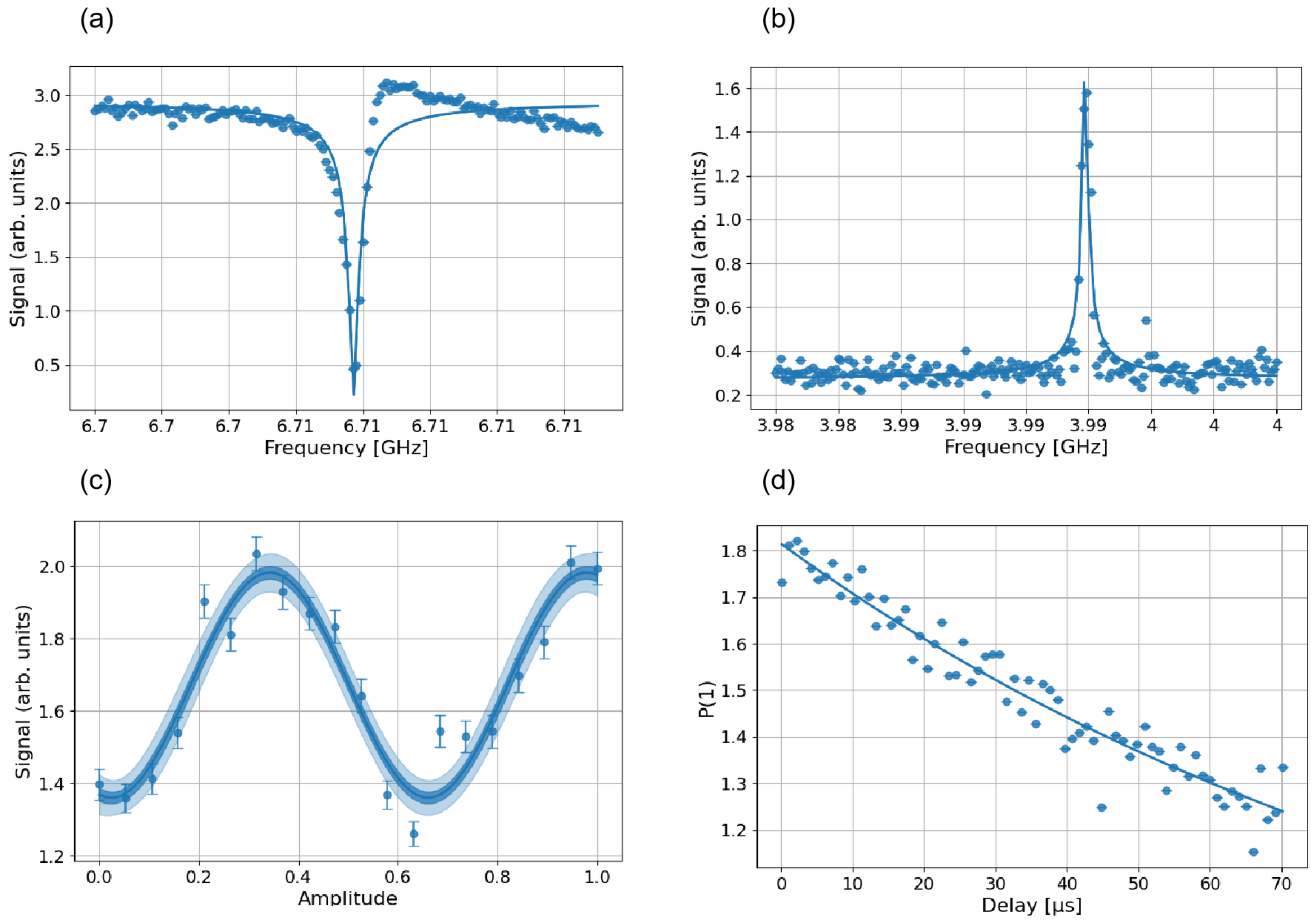

**Figure 8.** Subfigures (a) and (b) depict measurements of the readout cavity and qubit spectroscopy, respectively, showcasing the interaction between the quantum bit and its electromagnetic environment. Subfigure (c) illustrates qubit Rabi oscillations, revealing the dynamic behavior of the qubit under the influence of an external microwave field. Subfigure (d) presents measurements of the qubit's T1 relaxation time.

The S21 experiment is designed to evaluate the coupling characteristics between the qubit and the resonator. This is achieved by measuring the signal transmission efficiency between two ports, mediated by either the qubit or the microwave resonator. The procedure involves varying the frequency of the quantum bit and documenting the resultant amplitude and phase variations in the transmitted signal, the result is depicted in Fig. 8(a). In the Qubit Spectrum Experiment, the objective is to ascertain the resonance frequency of the qubit. This is conducted by applying microwaves at different frequencies and monitoring the qubit's response. The resonance frequency is identified at the point where the qubit exhibits its peak response. As illustrated in Fig. 8(b), the qubit probe frequency was swept around 3.992 GHz.The Rabi Oscillation Experiment aims to calibrate the amplitude of microwave pulses through the measurement of Rabi oscillations. By adjusting the microwave pulse width, the experiment observes the temporal evolution of the qubit's state, thereby determining the optimal pulse parameters for accurate qubit manipulation. This process is detailed in Fig. 8(c). T1 reflects the qubit's interaction strength with its surrounding environment and is essential for understanding the qubit's stability and operational lifespan, as shown in Fig. 8(d), the T1 of this qubit is 106us.

### C.Single-Qubit Performance Characterization Methods

1) **The single-shot experiment**

The single-shot measurement experiment, also referred to as the scatter experiment based on its adopted readout discrimination method, is a technique designed to determine the quantum state of a qubit in a single experimental trial. Unlike conventional approaches that rely on repeated measurements to extract averaged results, this method enables instantaneous and efficient readout of the qubit state, thereby minimizing measurement errors and enhancing operational efficiency. The experiment involves precise control over the timing and amplitude of measurement pulses to achieve single-shot state discrimination. The readout results are subsequently converted into classical signals for recording and analysis. Critical parameters influencing the experimental outcomes include the readout signal power, readout signal frequency, control signal power, and control signal frequency.

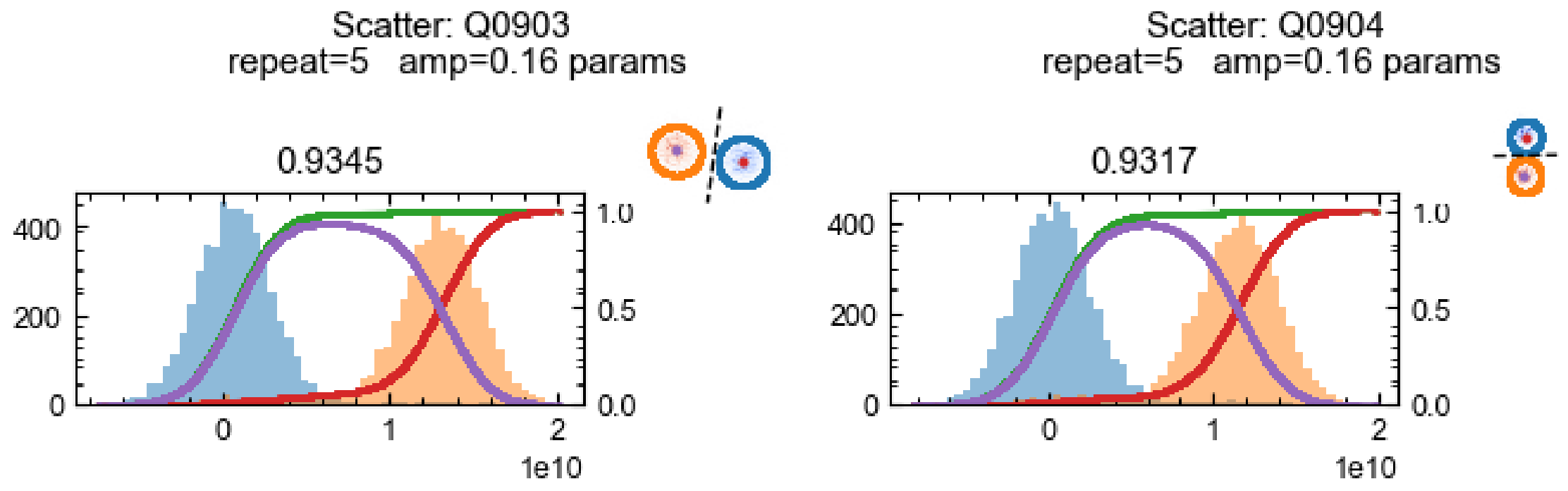

**Figure 9.** Single-Shot experimental results. (a) State projection histograms. (b) State clustering scatter plot on a two-dimensional plane.

The Single-Shot experiment enables accurate quantification of the signal distinguishability between the ground and excited states of a qubit, thereby determining the readout fidelity. Furthermore, the results provide the discrimination criteria (or resolution basis) for distinguishing the qubit's ground and excited states. This criterion serves as the fundamental reference for state discrimination in subsequent experiments. Thus, Single-Shot represents a critical intermediate process with milestone significance in quantum control and measurement workflows. These parameters must be meticulously optimized to ensure high-fidelity state discrimination and reliable quantum system characterization. As shown in Fig. 9, the single-shot readout fidelities of 93.45% (Q0903) and 93.17% (Q0904) demonstrate robust performance under the tested conditions.

2) **The Randomized Benchmarking experiment**

Randomized Benchmarking (RB) is one of the most widely employed quantum gate benchmarking methods. By performing and measuring sequences of randomly selected quantum gates, RB statistically evaluates the average fidelity of quantum gate operations. In RB experiments, the average fidelity of single-qubit gate operations can be determined by performing and measuring randomized gate sequences. By analyzing the fidelity decay curves for sequences of varying lengths, the rate of fidelity degradation is quantified, thereby evaluating the control quality and operational stability of the qubit.

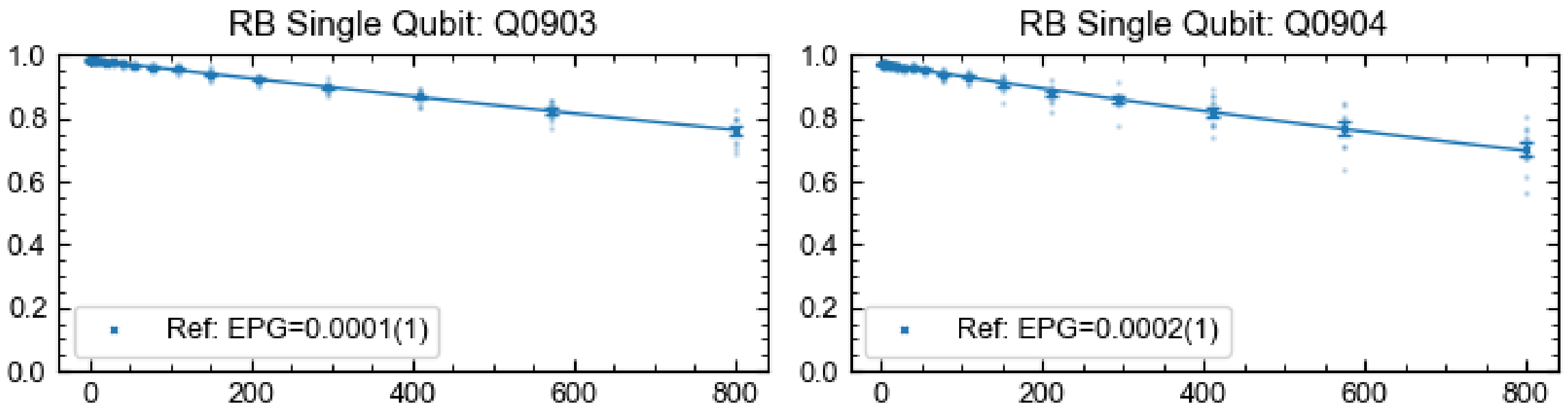

**Figure 10.** Single qubit gate benchmarking.

As illustrated in Fig. 10, the horizontal axis represents the number of quantum gates, while the vertical axis corresponds to the probability of the qubit occupying the excited state. As the number of gates increases, the excited-state probability decays exponentially, with the base of the exponential function representing the single-qubit gate fidelity. The calculated fidelity in the figure is 99.99(1) % of Q0903 and 99.98(1) % of Q0904.

### D. Two-Qubit Gate Performance Characterization Methods

By measuring changes in qubit states after applying a series of two-qubit gate sequences, the quality and performance of CZ gate operations can be assessed. For two-qubit gates, in the absence of external magnetic flux, phase accumulation inevitably persists due to intrinsic interactions. The role of the idle point is to apply counteracting magnetic flux to compensate for this phase accumulation. Therefore, prior to CZ gate calibration, the idle point must first be calibrated. Additionally, timing calibration is required due to inherent discrepancies between the Z-control and XY-control line configurations. Since the waveform of the CZ gate is typically short in duration, distortion is prone to occur on the chip, necessitating distortion calibration. As distortion calibration leverages the phase accumulation effect of the CZ gate—an effect highly sensitive to timing—distortion and timing calibrations must be performed in tandem. Subsequently, the amplitude (amp) and phase (phi) parameters of the CZ gate are calibrated using varying $N$ values for higher precision, enabling CZ-IRB experiments to determine the CZ gate fidelity. Next, leveraging the sensitivity of CZ gate fidelity to Z-signal parameters, the idle point is recalibrated. Finally, refined calibration of CZ gate parameters completes the high-fidelity CZ gate control protocol.

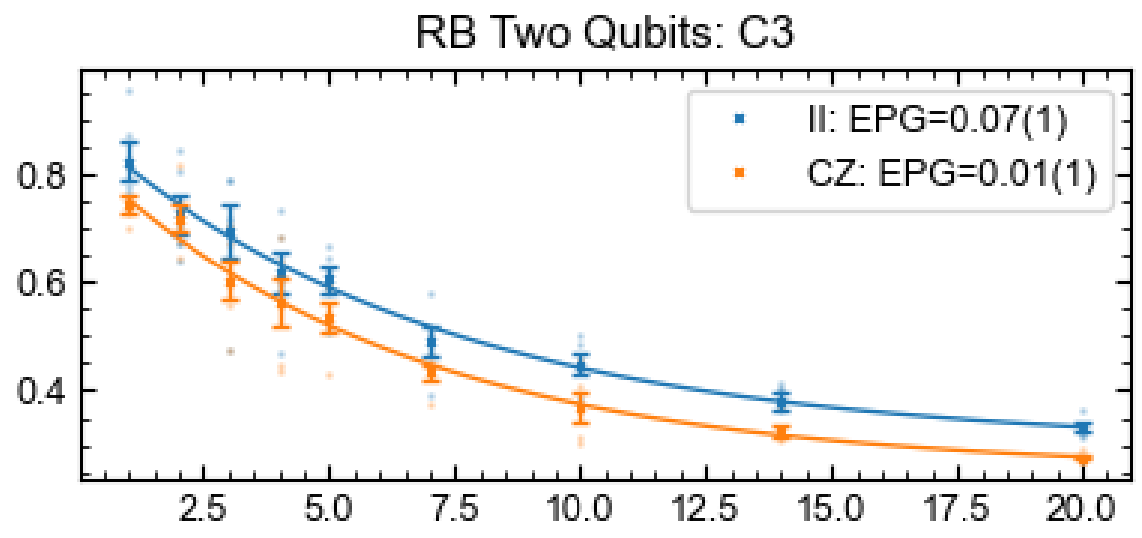

**Figure 11.** The CZ gate benchmarking with fidelity of 99%.

### E.Performance Test of Quantum Control and Measurement

As superconducting quantum chips progressively scale to medium-to-large sizes, the increasing volume of control and measurement tasks has rendered the qubit calibration process lengthy and time-intensive. This significantly impedes the iteration speed of quantum chips, emerging as a critical constraint on the advancement of quantum computing. Consequently, accelerating data processing, optimizing control and measurement workflows, and enhancing task throughput have become pressing priorities to achieve rapid and precise calibration for large-scale quantum chips.

In quantum control experiments, a task is divided into multiple steps based on experimental parameters. To mitigate the impact of random errors, each step is further subdivided into multiple shots during execution. This subdivision enhances experimental accuracy and reproducibility by obtaining statistical results through repeated shots, thereby reducing the influence of stochastic errors on outcomes. Here, a step is defined as the minimal unit for task generation and execution, while a shot represents a single physical realization of that task.

Under 500,000 shots, the new control and measurement board achieved an execution time of approximately 51 seconds, demonstrating an efficiency improvement by a factor of 4.94 compared to the legacy quantum control instrument's 252 seconds.

**F. System Latency Analysis**

In hybrid classical-quantum computing systems, latency is a critical performance metric that directly impacts the overall efficiency and scalability of quantum applications. To comprehensively evaluate the latency characteristics of the HI-HCQC system, we conducted a detailed analysis of the task execution pipeline, which encompasses both classical and quantum processing stages. The entire workflow can be decomposed into several sequential phases: task generation, data transmission, waveform sampling, signal playback and acquisition, signal demodulation, data upload, and data processing.

The classical preprocessing time includes waveform generation and initial data handling, which is largely dependent on the host CPU and software stack. However, the most significant variable in terms of latency lies in the data transmission phase between the classical host and the quantum control board. The theoretical lower bound of transferring a 2.4 MB waveform over a PCIe Gen3 x8 link with an effective bandwidth of 7.877 GB/s is approximately 305 μs. The measured transfer time is around 13 ms, indicating that software overhead, DMA setup cost, memory copy, driver latency, and OS scheduling dominate the practical data transfer latency. In practical measurements, however, the observed transmission time was around 13 ms, indicating the presence of software overhead, driver latency, and system scheduling factors.

A comparative test was performed against a conventional quantum control system (as used in quantum research institutes), which exhibited a data transmission latency of approximately 2.2 seconds with noticeable fluctuations. This stark contrast highlights a 169x speedup achieved by the HI-HCQC system in data transfer efficiency, underscoring the advantage of its PCIe-based architecture and optimized DMA mechanisms.

Another crucial contributor to total execution time is the repetition of measurement shots. Each shot includes a static reset period (typically set to 400 μs) to reinitialize the qubit to its ground state. For a standard sequence of 1,024 shots, the cumulative reset and acquisition time amounts to approximately 0.4 seconds. If the number of shots increases to 10,240, this duration extends to about 4 seconds. In such scenarios, the system becomes increasingly compute-bound, and further reductions in communication latency yield diminishing returns on overall application acceleration.

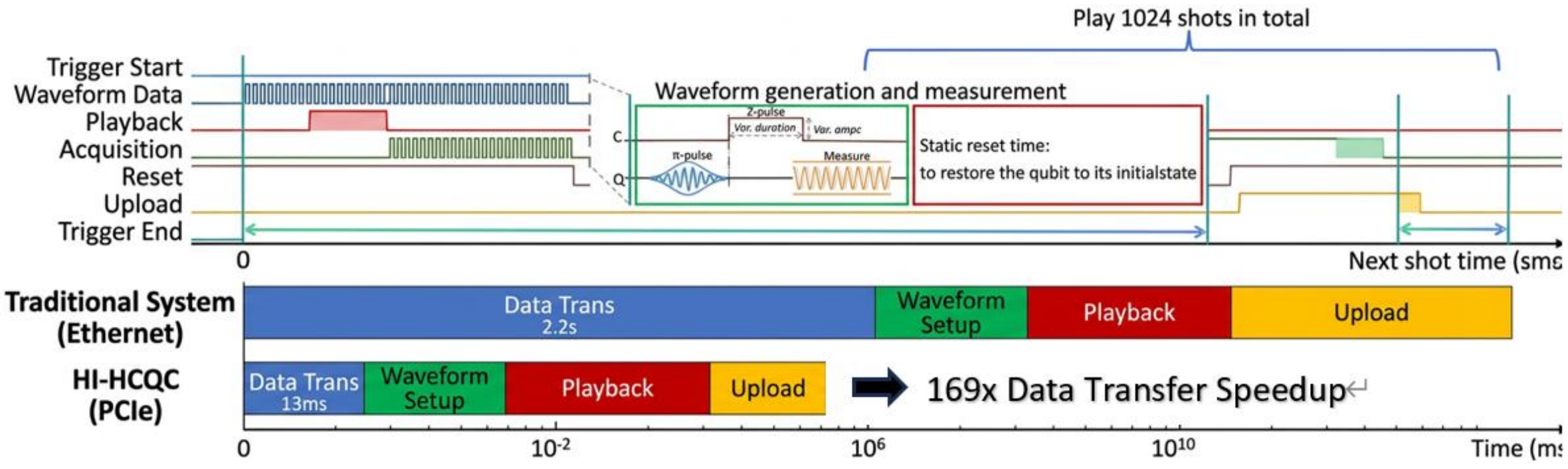

**Figure 12.** End-to-end latency breakdown comparison between the traditional Ethernet-based control system and the HI-HCQC PCIe-based control system.

This paper further conducts a systematic evaluation and comparison of the actual performance of the HI-HCQC control board in single-qubit gate, two-qubit gate, and multi-qubit quantum circuit operations. The performance acceptance experiments focus on the temporal characteristics of key quantum operations and system throughput capability. The reported latency includes waveform generation, data transfer, task scheduling, pulse playback, acquisition, demodulation, and result upload. It should not be interpreted as the physical duration of the quantum gate. By comparing with traditional control systems, the significant advantages of HI-HCQC in low latency and high throughput are verified. A quantitative comparison of key performance indicators is shown in Tab 3.

| Test Item | Traditional System | HI-HCQC | Speed-up Ratio | Performance Improvement |
|---|---|---|---|---|
| H Gate (Single-Qubit) Latency | 3.9309 s | 0.0166 s | 236.8 | Drastic reduction in communication and processing delay |
| CZ Gate (Two-Qubit) Latency | 4.3122 s | 0.0167 s | 258.2 | Benefiting from high-speed PCIe interface and optimized timing |
| GHZ Circuit Execution Latency | 4.2362 s | 0.0944 s | 44.875 | Improved multi-channel synchronization and overall scheduling efficiency |
| Quantum Gate Throughput | 0.4 | 128.3 | 320.75 | Optimization of hardware parallelism and task management mechanism |

**Table 3.** Comparison of Key Performance Indicators between HI-HCQC and Traditional Control Systems.

In the latency tests for single-qubit and two-qubit gates, HI-HCQC demonstrates excellent response performance. The latency measurements for the H gate (Hadamard gate) and the CZ gate (Controlled-Z gate) indicate that the delay in signal generation and feedback processing when the board executes these fundamental quantum gate operations is significantly lower than that of the traditional control system. This improvement primarily benefits from the high-speed PCIe interface and the optimized DMA data transfer mechanism adopted by HI-HCQC, which substantially reduces the communication overhead between the host and the FPGA, enabling faster transmission of control pulses and return of readout results. The low-latency characteristic is particularly crucial for quantum error correction protocols and dynamic circuit execution that require fast feedback, effectively reducing decoherence effects on quantum states during waiting periods and enhancing the execution fidelity of complex quantum algorithms.

Furthermore, in the runtime latency test for multi-qubit quantum circuits, we selected the GHZ state preparation circuit as a benchmark. When HI-HCQC runs this multi-qubit entanglement generation circuit, the overall timing control is tight, with high synchronization precision between pulses. Compared to the traditional control system, HI-HCQC significantly reduces the overall latency from instruction issuance to circuit completion, validating the effectiveness of its hardware architecture in coordinating multi-channel concurrent operations and maintaining global system synchronization. This performance is of great significance for scaling to large-scale quantum processors and executing multi-qubit gate sequences requiring precise timing coordination.

In terms of quantum gate throughput testing, HI-HCQC also demonstrates outstanding performance. The number of quantum gates that the board can schedule and execute per unit time far exceeds that of the traditional control system. The high throughput capability stems from its hardware parallel processing architecture and efficient task management mechanism, enabling the system to rapidly process quantum circuit descriptions containing a large number of gate operations and efficiently drive the signal generation and acquisition of the RF chain. This characteristic directly accelerates application workflows that require repeated execution of massive gate operations, such as quantum benchmarking and variational quantum algorithm iterations, providing hardware support for the rapid calibration and performance characterization of large-scale quantum chips.

These findings lead to three key conclusions:

Firstly, the number of shots significantly influences the execution time of quantum applications and must be carefully considered in the context of specific algorithms. For instance, the Bernstein–Vazirani algorithm may require only a single shot in the noiseless case, while variational algorithms such as VQE and randomized circuit sampling (RCS) often demand thousands to millions of shots to achieve statistical significance.

Secondly, when the number of shots is small, communication latency between classical and quantum components becomes a critical bottleneck. Optimizing data transfer mechanisms—as demonstrated by the HI-HCQC—can substantially improve system performance. Conversely, for shot-intensive applications, the system becomes primarily limited by quantum operation and reset times, thereby reducing the relative impact of communication improvements.

Thirdly, this latency analysis underscores the importance of a balanced co-design of classical and quantum components to achieve optimal performance in hybrid systems. The HI-HCQC represents a significant step forward in minimizing communication overhead and enhancing the scalability of practical quantum computing setups.

## Conclusion

In our research, we developed a system that leverages an RFSoC-based architecture to achieve efficient and precise control of quantum bits, along with the synthesis and measurement of microwave signals. This system is integrated with conventional computing resources, including servers and GPUs, via the PCIe interface. The modular design ensures high flexibility and adaptability, making it a versatile platform for quantum experiments.

Looking ahead, our objective is to enhance the system's scalability, enabling it to accommodate an increased number of qubits and support more complex experimental setups, thereby addressing the escalating demands of quantum computing research. We also plan to explore more sophisticated experimental control algorithms to refine the accuracy and efficiency of our experiments. Moreover, we are dedicated to enhancing the system's user-friendliness by streamlining the quantum programming and experimental setup processes, with the goal of broadening the community of researchers and engineers engaged in quantum computing.

**Acknowledgements**

This work is supported by National Key Research and Development Program of China (Grant No. 2023YFB4502500).

**References**

[1] Shor, Peter W. "Polynomial-time algorithms for prime factorization and discrete logarithms on a quantum computer." SIAM review 41.2 (1999): 303-332.
[2] Cao, Yudong, et al. "Quantum chemistry in the age of quantum computing." Chemical reviews 119.19 (2019): 10856-10915.
[3] Biamonte, Jacob, et al. "Quantum machine learning." Nature 549.7671 (2017): 195-202.
[4] Reagor, Matthew, et al. "Demonstration of universal parametric entangling gates on a multi-qubit lattice." *Science advances* 4.2 (2018): eaao3603.
[5] Cirac, Juan I., and Peter Zoller. "Quantum computations with cold trapped ions." *Physical review letters* 74.20 (1995): 4091.
[6] Zhong, Han-Sen, et al. "Quantum computational advantage using photons." *Science* 370.6523 (2020): 1460-1463.
[7] Blais, Alexandre, et al. "Quantum-information processing with circuit quantum electrodynamics." *Physical Review A* 75.3 (2007): 032329.
[8] Preskill, John. "Quantum computing in the NISQ era and beyond." *Quantum* 2 (2018): 79.
[9] “Xilinx RFSOC Website,” https://www.xilinx.com/products/silicon-devices/soc/rfsoc.html (2024).
[10] Minev, Zlatko. "Superconducting Qubits: Circuit Theory, Hamiltonian Analysis and Design Tools." *APS March Meeting Abstracts*. Vol. 2021. 2021.
[11] Werninghaus, Max, et al. "Leakage reduction in fast superconducting qubit gates via optimal control." *npj Quantum Information* 7.1 (2021): 14.
[12] Long, Junling, et al. "A universal quantum gate set for transmon qubits with strong ZZ interactions." *arXiv preprint arXiv:2103.12305* (2021).
[13] Bækkegaard, Thomas, et al. "Realization of efficient quantum gates with a superconducting qubit-qutrit circuit." *Scientific reports* 9.1 (2019): 13389.
[14] Vepsäläinen, Antti, et al. "Improving qubit coherence using closed-loop feedback." *Nature Communications* 13.1 (2022): 1932.
[15] Verjauw, J., et al. "Path toward manufacturable superconducting qubits with relaxation times exceeding 0.1 ms." *npj Quantum Information* 8.1 (2022): 93.
[16] Yap, Yung Szen, et al. "A Ku band pulsed electron paramagnetic resonance spectrometer using an arbitrary waveform generator for quantum control experiments at millikelvin temperatures." *Review of Scientific Instruments* 86.6 (2015).
[17] Riste, Diego, et al. "Detecting bit-flip errors in a logical qubit using stabilizer measurements." *Nature communications* 6.1 (2015): 6983.
[18] Raftery, J., et al. "Direct digital synthesis of microwave waveforms for quantum computing." *arXiv preprint arXiv:1703.00942* (2017).
[19] Kjaergaard, Morten, et al. "Superconducting qubits: Current state of play." *Annual Review of Condensed Matter Physics* 11 (2020): 369-395.

[20] Krantz, Philip, et al. "A quantum engineer's guide to superconducting qubits." *Applied physics reviews* 6.2 (2019).
[21] Kaufmann, Thomas, et al. "DAC-board based X-band EPR spectrometer with arbitrary waveform control." *Journal of Magnetic Resonance* 235 (2013): 95-108.
[22] Ryan, Colm A., et al. "Hardware for dynamic quantum computing." *Review of Scientific Instruments* 88.10 (2017).
[23] Salathé, Yves, et al. "Low-latency digital signal processing for feedback and feedforward in quantum computing and communication." *Physical Review Applied* 9.3 (2018): 034011.
[24] Lin, Jin, et al. "High performance and scalable AWG for superconducting quantum computing." *arXiv preprint arXiv:1806.03660* (2018).
[25] Sun, Lihua, et al. "Scalable self-adaptive synchronous triggering system in superconducting quantum computing." *IEEE Transactions on Nuclear Science* 67.9 (2020): 2148-2154.
[26] Kalfus, William D., et al. "High-fidelity control of superconducting qubits using direct microwave synthesis in higher Nyquist zones." *IEEE Transactions on Quantum Engineering* 1 (2020): 1-12.
[27] Xu, Yilun, et al. "QubiC: An open-source FPGA-based control and measurement system for superconducting quantum information processors." *IEEE Transactions on Quantum Engineering* 2 (2021): 1-11.
[28] Yang, Y., et al. "FPGA-based electronic system for the control and readout of superconducting qubit systems." *arXiv e-prints* (2021): arXiv-2110.
[29] Ryan, Colm A., et al. "Hardware for dynamic quantum computing." *Review of Scientific Instruments* 88.10 (2017).
[30] "Keysight Website,"https://www.keysight.com/us/en/solutions/emerging-technologies/quantum-solutions.html (2024).
[31] "Zurich Instruments Website," https://www.zhinst.com/americas/en (2024).
[32] "Quantum Machines Website," https://www.quantum-machines.co/ (2024).
[33] Xu, Yilun, et al. "Qubic 2.0: An extensible open-source qubit control system capable of mid-circuit measurement and feed-forward." *arXiv preprint arXiv:2309.10333* (2023).
[34] Tholén, Mats O., et al. "Measurement and control of a superconducting quantum processor with a fully integrated radio-frequency system on a chip." *Review of Scientific Instruments* 93.10 (2022).
[35] Stefanazzi, Leandro, et al. "The QICK (Quantum Instrumentation Control Kit): Readout and control for qubits and detectors." *Review of Scientific Instruments* 93.4 (2022).
[36] Singhal, Ujjawal, et al. "SQ-CARS: A Scalable Quantum Control and Readout System." *IEEE Transactions on Instrumentation and Measurement* (2023).
[37] Park, Kun Hee, et al. "ICARUS-Q: Integrated control and readout unit for scalable quantum processors." *Review of Scientific Instruments* 93.10 (2022).
[38] Pauka, S. J., et al. "A cryogenic CMOS chip for generating control signals for multiple qubits." *Nature Electronics* 4.1 (2021): 64-70.
[39] Hornibrook, J. M., et al. "Frequency multiplexing for readout of spin qubits." *Applied Physics Letters* 104.10 (2014).
[40] "Python productivitye for zynq," https://pynq.io (2024).
[41] Alexander, Thomas, et al. "Qiskit pulse: programming quantum computers through the cloud with pulses." Quantum Science and Technology 5.4 (2020): 044006.
[42] Naghiloo, Mahdi. "Introduction to experimental quantum measurement with superconducting qubits." *arXiv preprint arXiv:1904.09291* (2019).
[43] Krantz, Philip, et al. "A quantum engineer's guide to superconducting qubits." *Applied physics reviews* 6.2 (2019).
[44] Kim, Y., Eddins, A., Anand, S. *et al.* Evidence for the utility of quantum computing before fault tolerance. *Nature* 618, 500–505 (2023).
[45] Bilmes, A., Megrant, A., Klimov, P. *et al.* Resolving the positions of defects in superconducting quantum bits. *Sci Rep* 10, 3090 (2020).

**Appendix A**

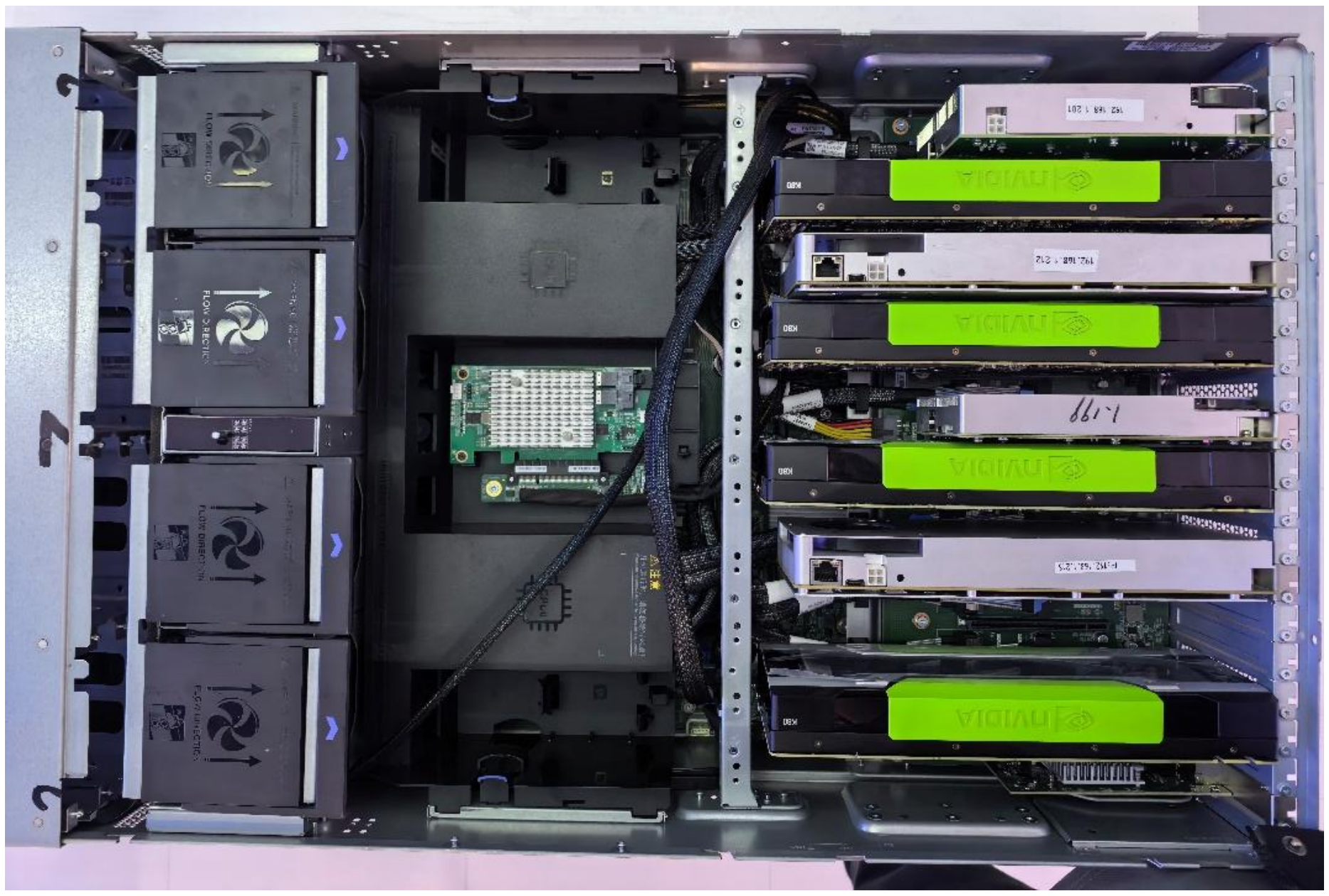

**Figure A-1**

This diagram illustrates the internal hardware layout of a server chassis designed for classical-quantum heterogeneous collaborative computing scenarios. Adopting layered heat dissipation and modular hardware architecture, the chassis is divided into three sections from left to right: the thermal module layer, central main control layer, and acceleration & measurement-control hardware layer.

The left thermal module consists of four independent modular cooling fan units, which deliver forced air cooling downward uniformly to all internal hardware components of the server. This ensures long-term stable operation of high-density hardware including the CPU motherboard, GPU accelerator cards, and quantum measurement & control circuit boards.

The central main control section houses the core server CPU motherboard, fitted with two Intel processors (12 cores, 24 threads, 2.0 GHz). It is equipped with four 16 GB DDR5 ECC REG memory modules, two 480 GB 2.5-inch enterprise-grade SATA SSD system drives, an SAS3008 RAID controller (supporting RAID 0.1), and two Gigabit Ethernet adapters. This section undertakes core tasks such as classical data processing, quantum circuit compilation, and post-data processing. Four 2700 W redundant power supplies are deployed within the entire chassis to deliver stable power supply for high-power hardware components.

Expansion slots on the right side accommodate four NVIDIA Tesla K80 GPU accelerator cards in sequence, supplying parallel computing power for quantum simulation, classical post-processing and other workloads. Mounted vertically in high-density stacks within these slots are three types of measurement-control boards: XY and Z boards, with four XY short boards and four Z long boards respectively. Dense high-speed signal pins and interfaces are distributed across these boards to transmit high-speed measurement and control signals, serving as bridges for signal relay and timing control between classical computing units and quantum qubit hardware.

This architecture achieves tightly coupled integrated deployment combining general CPU main control, classical GPU acceleration, and low-latency measurement & control of dedicated quantum hardware